\documentclass[twocolumn,prl,10pt,showpacs,superscriptaddress]{revtex4-1}
\usepackage{graphicx}  
\usepackage{amssymb} 
\usepackage{natbib}
\usepackage{amsmath}
\usepackage{enumerate}
\usepackage{textcomp}
\usepackage{mathtools}
\usepackage{epsfig}
\usepackage[colorlinks,linkcolor=blue,citecolor=blue,urlcolor=blue]{hyperref}
\usepackage{color}
\usepackage{graphicx}
\usepackage[dvipsnames]{xcolor}
\usepackage{tabularx}
\usepackage{soul}
\usepackage[mathscr]{euscript}
\usepackage{amssymb}
\usepackage[normalem]{ulem}
\usepackage{mathrsfs}
\usepackage{amsmath}
\usepackage{color}
\usepackage{bbold}
\usepackage{comment}
\usepackage{bm}
\usepackage{braket}
\usepackage{dsfont}
\usepackage{textcomp}
\usepackage{lipsum}
\usepackage{amsmath,amssymb}
\newcommand{\pt}{\partial}
\newcommand{\mb}{\mathbf}
\newcommand{\mc}{\mathcal}

\newcommand{\Fext}{F_{\scalebox{0.6}{ext}}}

\newcommand{\Hext}{H_{\scalebox{0.6}{ext}}}
\newcommand{\hext}{h_{\scalebox{0.6}{ext}}}

\newcommand{\Xmin}{X_{\scalebox{0.6}{min}}}

\newcommand{\Vmax}{V_{\scalebox{0.6}{max}}}

\begin{document}

\title{Quantum Depinning of a Magnetic Skyrmion}

\author{Christina Psaroudaki}
\affiliation{Department of Physics, University of Basel, Klingelbergstrasse 82, 4056 Basel, Switzerland}
\author{Daniel Loss}
\affiliation{Department of Physics, University of Basel, Klingelbergstrasse 82, 4056 Basel, Switzerland}

\date{\today}
\begin{abstract}
We investigate the quantum depinning of a weakly driven skyrmion out of an impurity potential in a mesoscopic magnetic insulator. For small barrier height, the Magnus force dynamics dominates over the inertial term, and the problem is reduced to a massless charged particle in a strong magnetic field. The universal form of the WKB exponent, the rate of tunneling, and the crossover temperature between thermal and quantum tunneling are provided, independently of the detailed form of the pinning potential. The results are discussed in terms of macroscopic parameters of the insulator Cu$_2$OSeO$_3$ and various skyrmion radii. We demonstrate that small enough magnetic skyrmions, with a radius of $\sim$10 lattice sites, consisting of some thousands of spins, can behave as quantum objects at low temperatures in the millikelvin regime.  
  \end{abstract}

\maketitle
Magnetic systems have been theoretically predicted
\cite{Hemmen86,Stamp91,Stamp92,Chudnovsky92,Loss92,Tatara93,Braun96,Braun97,LossBook} and experimentally verified 
\cite{Awschalom90,Awschalom92,Barbara95,Thomas96,Wernsdorfer99,Brooke01,Zarzuela12} to be good candidates for the observation of macroscopic quantum tunneling events and quantum to classical phase transitions \cite{Owerre15}. In such systems, a large number of elementary magnetic moments display quantum behavior, as they may coherently tunnel from a metastable configuration to a more stable state. A particle-like configuration of the classical magnetization field, supports a collective mode of position that tunnels out of the local minimum through a potential barrier into the classically forbidden region. 

Among the various magnetic solitons, skyrmions are in the focus of current research because they appear as attractive candidates for future spintronic devices \cite{Finocchio16,Fert17}. Skyrmions are spatially localized two-dimensional (2D) topological magnetic textures in a magnetic material, which can be either metallic \cite{Muehlbauer09}, a multiferroic insulator \cite{Seki12}, or ultrathin metal film on heavy-element substrates \cite{Heinze11}. Typically they are classical objects with a size of the order of $50$ nm and a dynamics that is governed by the Landau-Lifshitz-Gilbert (LLG) equation \cite{lifshitzBK80,gilbertTM04}, 
although small-size skyrmions of $1$ nm (a few lattice constants) have been recently observed \cite{Wiesendanger16}, inspiring studies on the quantum properties of skyrmions \cite{Lin13,Takashima16,Diaz16,Psaroudaki17,Psaroudaki18,Derras-Chouk18,Psaroudaki19,Sotnikov18,Lohani19,Vlasov19}. 

In this Letter, we study the quantum depinning of a magnetic skyrmion out of a potential created by an atomic defect in a magnetic insulator, which removes the very strong dissipation from itinerant electrons. We consider the application of a magnetic field gradient, which tilts the pinning potential, and lowers the barrier height. The skyrmion then escapes out of its metastable state into the classically forbidden region along the direction of the magnetic field gradient. In the limit of small barrier height, inertial terms can be neglected, and the skyrmion dynamics is governed by the Magnus force. In this respect, the skyrmion dynamics resembles the Hall type dynamics of a vortex in high-$T_c$  superconductors \cite{Gorokhov98,Blatter94}, or a charged spin texture in Quantum Hall systems \cite{Abolfath01,Kyriakidis99}. Nevertheless, skyrmions in insulating ferromagnets provide the unique setup to study tunneling events in situations when the Magnus force, related to the \textit{topological character} of the particle, is the dominant contribution over the intertial term. The exact temperature-independent WKB exponent depends on the pinning potential width, but not on its depth in contrast to the result for a massive magnetic particle. We provide explicit expressions for the tunnel frequency, the tunneling rate, and the crossover temperature between quantum tunneling and thermal activation, for arbitrary width and height of the pinning potential. We give estimates of these quantities for the magnetic insulator Cu$_2$OSeO$_3$, and find that skyrmions, under certain specified circumstances, can exhibit macroscopic quantum behavior. 

To study the macroscopic tunneling of a magnetic texture $\mb{m}$ from a defect pinning center, we employ the imaginary time formulation for path integrals with a Euclidean action written in the form,
\begin{align}
\mc{S}_E = N_A \int_{0}^{\beta} d\tau \left( i S d^2\int d \mb{r} \dot{\Phi} (1-\Pi) +  \mc{H}/J_0 \right) \,,
\label{EuclAction}
\end{align}
where the magnetization density at position $\mb{r}$ is represented in polar coordinates, $\mb{m} (\mb{r})=[\sin \Theta \cos \Phi,\sin \Theta \sin \Phi,\cos \Theta]$, $\Pi = \cos \Theta$, $S$ is the total spin, and $N_A$ the number of layers. The magnetic Hamiltonian $\mc{H} = J_0 \int d\mb{r} \mc{F}(\mb{m})$ reads,
\begin{align}
\mc{F}(\mb{m})=  \sum_{i=x,y} (\frac{\pt \mb{m}}{\pt r_i})^2 + \mb{m} \cdot \nabla \times \mb{m} -\kappa m_z^2 -h m_z  \,.
\label{FreeEnergy}
\end{align}
The exhange coupling $J_0$ sets the energy scale, while $\kappa = K J_0/D_0^2$ and $h=g \mu_B H J_0/D_0^2$ are dimensionless and denote the strength of anisotropy and uniform magnetic field, respectively. $D_0$ and $K$ denote the Dzyaloshinskii-Moryia (DM) and anisotropy coupling respectively in units of energy, and $H$ the external magnetic field in units of T. Imaginary time $\tau$ and space $\mb{r}$ variables are given in reduced units. Physical units are restored as $\mb{r}' = \mb{r}  \alpha d$ and $\tau' = \tau /J_0$, where $d=J_0/D_0$, and $\alpha$ is the lattice constant. Also, we set $\hbar = 1$. In $\mc{F}(\mb{m})$, Eq.~\eqref{FreeEnergy}, we ignore dipolar magnetic interactions, which are usually weaker than the DM coupling and would stabilize skyrmions of much larger size than considered here. The functional Eq.~\eqref{FreeEnergy} supports a stable skyrmionic solution, described by $\Phi=\phi+ \pi/2$ and the approximate function $\Theta (\rho)= 2 \tan^{-1}[(\lambda/\rho) e^{-(\rho -\lambda)/\rho_0}]$, with $(\rho,\phi)$ the polar coordinate system, $\rho_0 = \sqrt{2/(2 \kappa+h)}$, while $\lambda$, which we obtain numerically from the Euler-Lagrange equation of the stationary skyrmion, is the skyrmion radius \cite{Psaroudaki17}. Magnetic skyrmions are characterized by a finite topological charge $Q$,
\begin{align}
Q=\frac{1}{4 \pi} \int d\mb{r}~ \mb{m} \cdot (\pt_x \mb{m} \times \pt_y \mb{m}) \,,
\end{align}
which denotes the mapping from the 2D magnetic system in real space into the 3D spin space. 

\begin{figure}[t]
\centering
\includegraphics[width=1\linewidth]{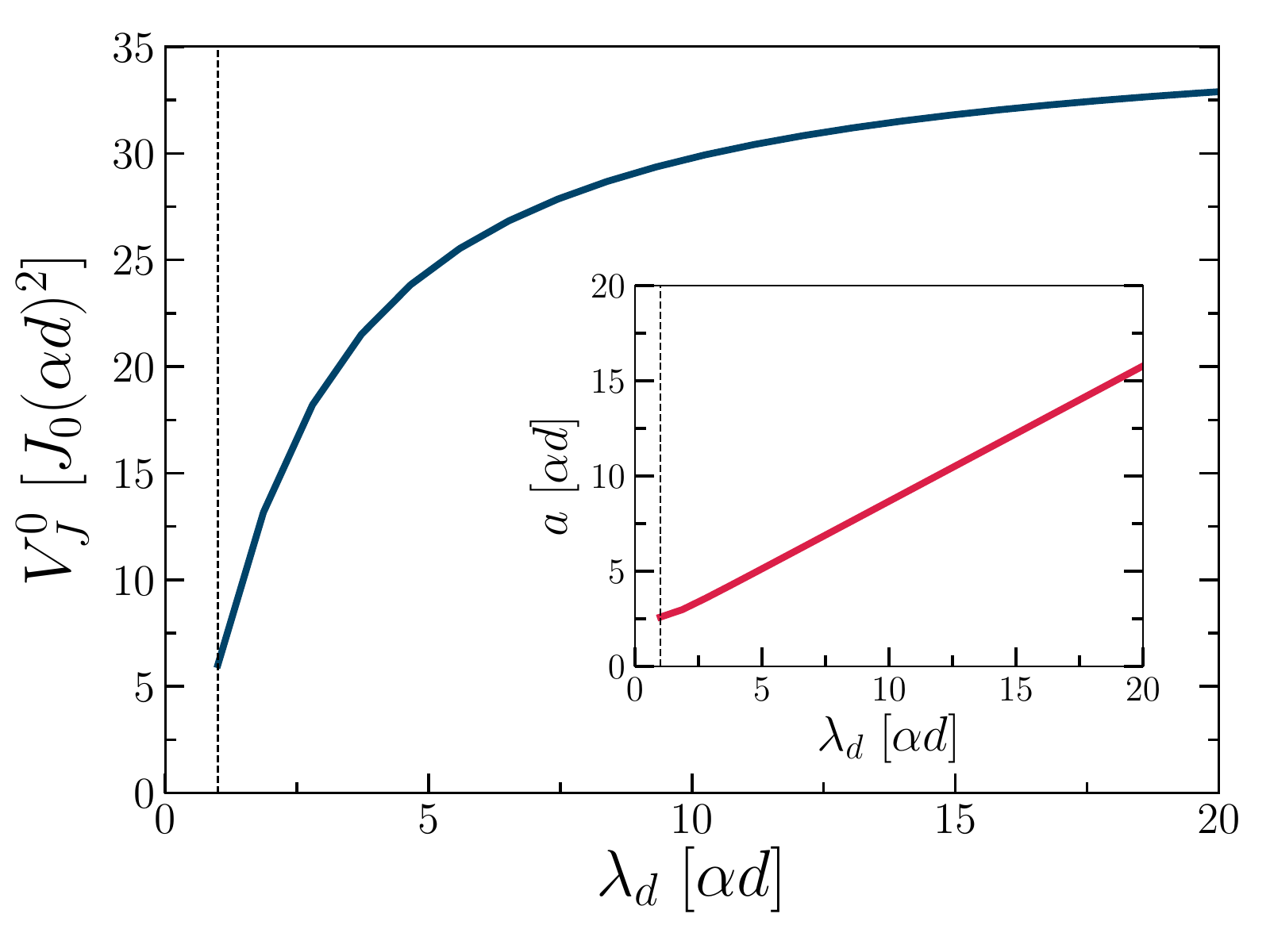}
  \caption{Parameters $V_0,a$ of the pinning potential $V_p(r_0)$ of Eq.~\eqref{Pinning}, as a function of the defect size $\lambda_d$, with $\kappa = 0.463$ and $h=0.25$, for a skyrmion of radius $\lambda = 1.86 \alpha d$, with $\alpha$ the lattice constant. Vertical dashed line indicates the value $\lambda_d =\alpha d$. } 
\label{fig:aJVJ0} 
\end{figure}

The presence of a crystal defect at $\mb{r} = 0$ alters the exchange and DM couplings as $J/J_0= 1- J' e^{-\rho/\lambda_d}$ and $D/D_0= 1 - D' d e^{-\rho/\lambda_d}$, respectively \cite{Lin13}, with $J', D'$ being the strength, and $\lambda_d$ the size of the defect. On the classical level, the interactions of skyrmions with atomic defects crucially affect their mobility \cite{Lin13b,Muller15,Stosic17,Diaz18,Fernandes18,Liu13}. $J'$ and $D'$ are perturbations and the distortion of the skyrmion profile is weak. The resulting pinning potential $V_p$ as a function of the distance $r_0$ between the center of the defect and the center of the skyrmion can be approximated by the function 
\begin{align}
 V_p(r_0) = - \frac{1}{J_0}\frac{V_0(\lambda_d)}{r_0^2 +a(\lambda_d)^2} \,,
 \label{Pinning}
 \end{align}
where $V_0(\lambda_d)= c_0 N_A V_J^{0}(\lambda_d)$, $c_0=(1-d/d') J' /J_0$ and $d'=J'/D'$ \cite{Supp}. We take $V_0(\lambda_d)>0$, in order for the skyrmion to experience an attractive potential. The behavior of $a(\lambda_d)$ and $V_J(\lambda_d)$ is summarized in Fig.~\ref{fig:aJVJ0}. 

To describe a skyrmion escaping from the potential well, we need to employ a description which isolates the center of mass of the skyrmion, $\mb{R}(\tau)$, from all other degrees of freedom. This is achieved by integrating out the magnon degrees of freedom, within a quantum field theory method which makes use of the Faddeev-Popov techniques for collective coordinates \cite{Psaroudaki17}. Then one finds that the Euclidean action of Eq.~\eqref{EuclAction} takes the form,
\begin{align}
\mc{S}_{E} = \int_{0}^{\beta} d\tau [ i\tilde{Q} (\dot{\mc{X}} \mc{Y}-\dot{\mc{Y}} \mc{X}) +\frac{1}{2}\mc{M} \dot{\mb{R}}^2+U(\mc{X},\mc{Y}) ] \,,
\label{Action}
\end{align}
with $\mb{R} = (\mc{X},\mc{Y})$, and $\tilde{Q}=2 \pi N_A S Q$. Here $U(\mc{X},\mc{Y}) = V_p(\sqrt{\mc{X}^2+\mc{Y}^2}) -\Fext \mc{X}$, where $V_p$ being the pinning potential of Eq.~\eqref{Pinning}, and $\Fext$ is a linear force acting on the skyrmion collective coordinate equal to $\Fext = \hext \pt/\pt \mc{X} \int d \mb{r}~ x m_z(\mb{r}-\mb{R})$, as the result of an applied out-of-plane magnetic field gradient. We introduce $\hext =   g\mu_B N_A \Hext J_0^2/D_0^3$ and $\Hext$ is measured in T. The first term in \eqref{Action} is a Magnus force acting on the skyrmion proportional to the topological number \cite{Stone96}. Here, $\mc{M}$ denotes the effective mass which arises from the skyrmion-magnon bath coupling in the presence of a pinning potential \cite{Psaroudaki17}, (for  explicit expressions for $\mc{M}$, see Ref.~\onlinecite{Psaroudaki17} and, in particular, Eq.~(32) therein). A non-negligible mass term gives rise to oscillatory modes in the real-time dynamics of the skyrmion \cite{Makhfudz12,Buttner15}, which performs a cyclotron rotation of frequency $\propto \tilde{Q}/\mc{M}$.  

For small values of the magnetic field $\Fext$, the skyrmion is trapped at its minimum position. As the field grows, the barrier \eqref{Pinning} is lowered and the skyrmion eventually gets depinned at the coercive force $F_c$. However, even for $\Fext<F_c$, the position of the skyrmion at the pinning center becomes metastable and can tunnel out of the local minimum, as long as $0<\epsilon \equiv 1-\Fext/F_c \ll 1$ \cite{Stamp91,Braun97} . The coercive force is given by $F_c = V'(R_i)$, where $R_i$ is the inflection point close to a local minimum, calculated by requiring $V_p''(R_i)=0$, $V_p'(R_i)>0$ and $V_p^{(3)}(R_i)<0$. For the potential of Eq.~\eqref{Pinning}, we find $R_i = a/\sqrt{3}$. 

Provided that effective mass in dimensionless units is $\mc{M} \propto N_A S U_0^2$, where $U_0$ is the height of the barrier \cite{Psaroudaki17}, and motivated by the fact that the optimum condition for the observability of tunneling events is when the potential barrier is small and narrow, it is convenient to separate the fast cyclotron rotation of frequency, $\tilde{Q}/\mc{M} \gg 1$, from the slow motion of the guiding center \cite{Dmitriev95}. This is achieved by considering the real time Lagrangian $L$, obtained upon replacing imaginary time $\tau$ with real time $t =-i \tau$ in the imaginary time Lagrangian $\mc{L}$, with $\mc{S}_E = \int_0^{\beta} d\tau \mc{L}$ and $\mc{S}_E$ given in Eq.~\eqref{Action}. The Hamiltonian $H$ that corresponds to $L$ is given by
\begin{align}
H= \frac{1}{2\mc{M}} [(P_x + \tilde{Q} \mc{Y})^2 +(P_y - \tilde{Q} \mc{X})^2] + U(\mc{X},\mc{Y})\,,
\label{Hamiltonian}
\end{align}
with $P_x=\mc{M} \dot{\mc{X}} -\tilde{Q} \mc{Y}$ and $P_y=\mc{M} \dot{\mc{Y}}+\tilde{Q} \mc{X}$. Following Refs.~\cite{Dmitriev95,Gorokhov98}, instead of the original coordinates $\mc{X},\mc{Y}$ and conjugated momenta $P_x,P_y$, we define new operators $X,Y$ and $\Pi_x,\Pi_y$ as 
$X= P_y/\tilde{Q} +\mc{X} $, $Y =- P_x/ \tilde{Q} +\mc{Y}$, $\Pi_x =  - P_y/\tilde{Q} +\mc{X}$, and $ \Pi_y =  P_x/\tilde{Q} +\mc{Y}$, with $[X,Y]=-2i/\tilde{Q}=[\Pi_y,\Pi_x]$, while all other commutators vanish. These new operators form a complete set of canonical variables which can be used instead of the original coordinates. In the $X, \Pi_y$ representation, $\Pi_x=(2 i/\tilde{Q})\pt /\pt \Pi_y$, $Y=(2 i/\tilde{Q})\pt /\pt X$, and the Hamiltonian \eqref{Hamiltonian} equals
\begin{align}
H &= H_f + \int d\mb{k} U(k_x,k_y)e^{i \frac{k_x}{2} (X+\Pi_x) + i \frac{k_y}{2} (Y+\Pi_y)} \,, 
\end{align}
\begin{figure}[t]
\centering
\includegraphics[width=1\linewidth]{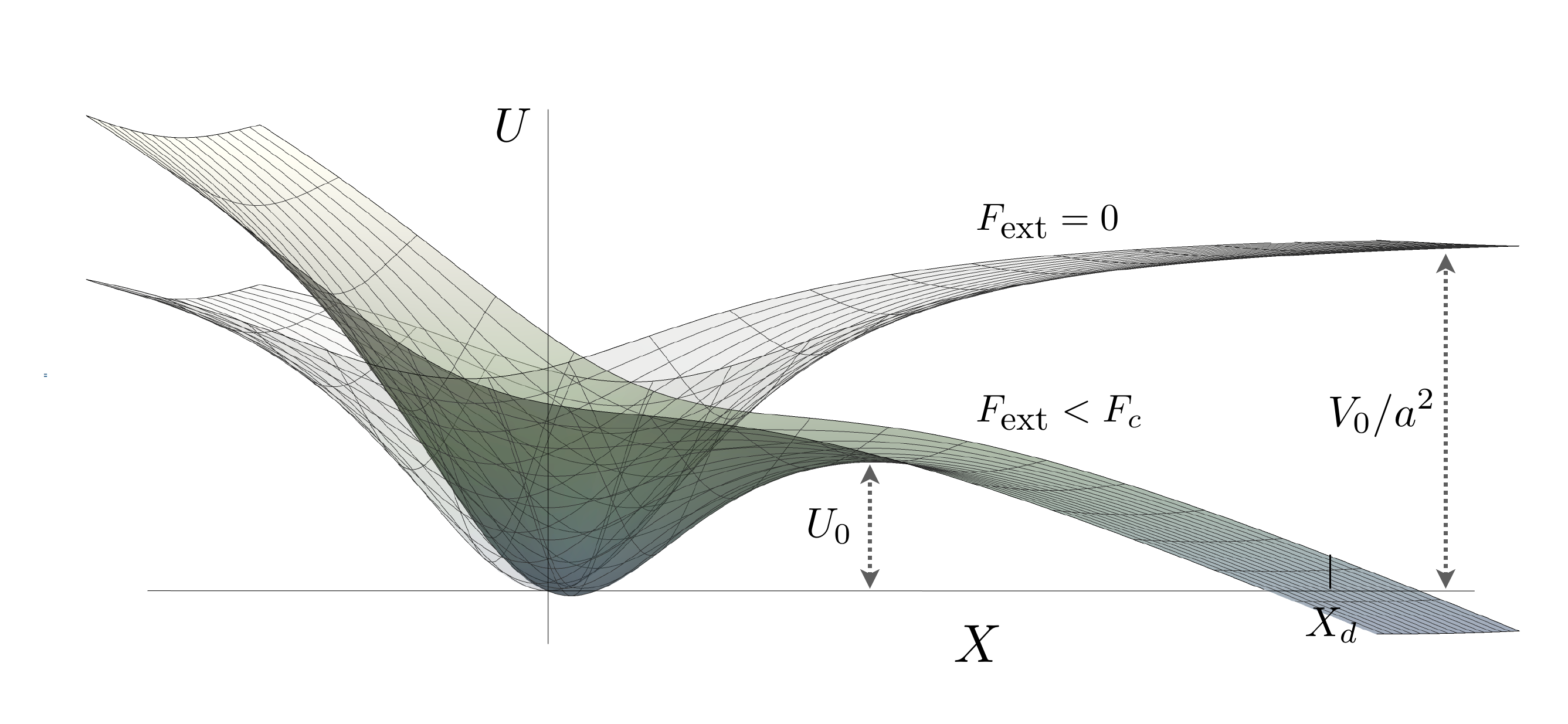}
  \caption{Schematic illustration of the pinning potential $U$ as function of position, for two values of the external field, $F_{\mbox{\scriptsize ext}}=0$, and $0< F_{\mbox{\scriptsize ext}}< F_c$. $U_0$ denotes the height of the potential barrier and $X_d$ the position of the turning point. } 
\label{fig:Pot_Tun} 
\end{figure}

where the fast part of the skyrmion Hamiltonian is expressed in terms of $H_f=-(1/2M)\pt^2/\pt \Pi_y^2 +  (1/2) M \omega_c^2 \Pi_y^2 $, and corresponds to a harmonic oscillator with frequency $\omega_c =2\tilde{Q}/\mc{M}$ and mass $M=\mc{M}/4$. The ground state $\vert \Psi_0 \rangle$ of $H_f$, $\langle \Pi_y \vert \Psi_0 \rangle=(2 \pi/\tilde{Q})^{-1/4}e^{-i \tilde{Q}\Pi_y^2/4}$, describes the cyclotron motion of the skyrmion at the zero Landau level with ground state energy equal to $\omega_c/2$. By averaging over the fast rotation, $\langle \Psi_0 \vert H \vert \Psi_0 \rangle$, and taking the zero Landau level as a reference point for energy, $\mc{H} = \langle \Psi_0 \vert H \vert \Psi_0 \rangle - \omega_c/2$, we obtain $\mc{H} = U(X, Y)$, with $[X,Y] = -i/2 \tilde{Q}$. This approximation holds as long as $l \ll \alpha$, where $l$ is the magnetic length $l = \alpha d \tilde{Q}^{-1/2}$, while in this limit $[X,Y] \rightarrow 0$. 
\begin{figure*}[t] 
\includegraphics[width=1\textwidth]{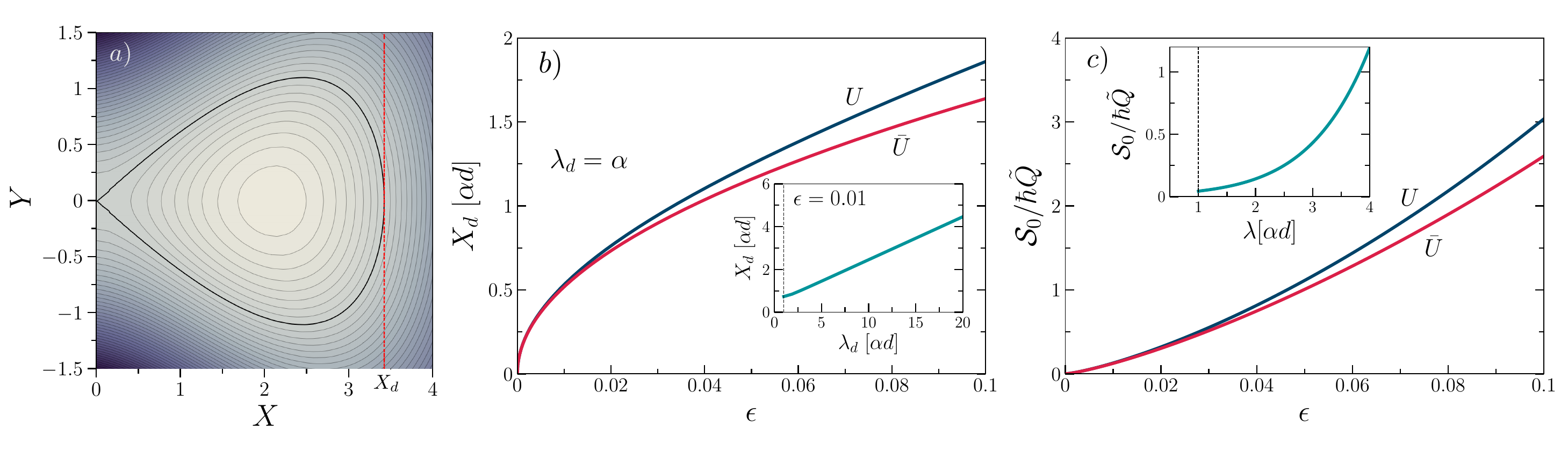}
\caption{(a) Equipotential lines of the potential landscape $U(X,iY)$, for a skyrmion of radius $\lambda = 1.86 \alpha d$, trapped by a pinning center of radius $\lambda_d=2.5 \lambda$, and $\epsilon=0.096$. The instanton trajectory describing the tunneling of the skyrmion, is depicted by the black solid line, while the red dashed line signals the position of the turning point $X_d$. (b) $X_d$ as a function of $\epsilon$, calculated using the full potential $U$ as well as the expanded potential $\bar U$ of Eq.~\eqref{ExpandPot}, for $\lambda= 1.86 \alpha d$ and $\lambda_d = \alpha d$. The expansion is valid for $\epsilon \lesssim 0.05$. The inset illustrates the linear dependence of $X_d$ on the defect size, for a choice of $\epsilon=0.01$. (c) The tunneling exponent $\mc {S}_0$ as a function of $\epsilon$ for $\lambda= 1.86 \alpha d$ and $\lambda_d = \alpha d$, while the inset depicts the dependence of $\mc{S}_0$ on the skyrmion size. }
\label{fig:Merge}
\end{figure*}   

With these preparations, the problem is reduced to a problem equivalent to that of the motion of a massless charged particle in a strong magnetic field, with an action of the form
\begin{align}
\mc{S}_E =\int_0^{\beta} d\tau \left[  i\tilde{Q} (\dot{X} Y- \dot{Y} X)+ U(X,Y) \right]\,,
\label{LagrRT}
\end{align}
and a saddle point solution which is in general complex. The criterion for the applicability of the WKB method for the action Eq.~\eqref{LagrRT} is the same as the 1D massive case \cite{Gorokhov98}. We introduce a normalized potential of the form $U(X,Y) = V_p(X,Y) - \Fext X$, with $V_p (X,Y) = V_0(\lambda_d) [1/a(\lambda_d)^2 -1/(X^2+Y^2+a(\lambda_d)^2)]$, and $V_0, a$ as in \eqref{Pinning}. We further consider the potential in shifted coordinates $U(X,Y) \rightarrow U(X+\Xmin,Y) - U(\Xmin,0)$, where $\Xmin$ is defined as $\pt U(X,0)/\pt X \vert_{X=\Xmin} = 0$. In Fig.~\ref{fig:Pot_Tun} we plot the potential energy $U$ for $\Fext = 0$ and $0< \Fext < F_c$. The analysis is significantly simplified if we expand around the inflection point $X_i$, defined as $\pt^2 U(X,Y)/\pt X^2 \vert_{X=X_i,Y=0} = 0$. The resulting expression is,
\begin{align}
\bar{U}(X,Y) \simeq  \Vmax^y \frac{Y^2}{Y_d^2}(1- \frac{X}{Y_d}) + \Vmax^x \frac{X^2}{X_d^2}(1- \frac{X}{X_d}) \,.
\label{ExpandPot}
\end{align}

We also introduce $Y_d = c_1/c_2$, $X_d=c_4/c_3$, $\Vmax^y = c_1^3/c_2^2$, and $\Vmax^x = c_4^3/c_3^2$, where 
$c_1 = (1/2) V_{(0,2)} + [\epsilon F_c V_{(1,2)}^2/(-2 V_{(3,0)})]^{1/2}$, $c_2 =- (1/2) V_{(1,2)}$, $c_3=- (1/6)V_{(3,0)}$, and $c_4 = [-(1/2)\epsilon F_c V_{(3,0)}]^{1/2}$. Derivatives are denoted as $V_{(i,j)} = V_p^{(i,j)}(X_i,0)$, while $ V_p^{(3,0)}(X_i,0),~V_p^{(1,2)}(X_i,0)<0$. For the particular choice of the pinning potential, the parameters simplify as $c_1= c (1/\sqrt{3}+2 \sqrt{\epsilon}/3)$, $c_2 = c/a$, $c_3=c_2/2$, $c_4= c\sqrt{\epsilon}$, $X_d= 2 a \sqrt{\epsilon}$, and $\Vmax^x=4 a^2 c \epsilon^{3/2}$ with $c=9\sqrt{3}V_0/J_0 16 a^4$.   

To study the imaginary time trajectories and obtain a real problem from the action \eqref{LagrRT}, we perform the additional transformation $Y \rightarrow iY$, provided that the condition $\mbox{Im}[U(X,iY)]=0$ holds. The instanton trajectories $(X_I,Y_I)$ are the classical solutions of the equations of motion in Euclidean time, $2 \tilde{Q} \dot{Y_I} +\pt U/\pt X_I = 0$ and $-2 \tilde{Q} \dot{X_I} +\pt U/\pt Y_I = 0$. By integrating the first (second) equation with respect to $X_I$ ($Y_I$) we arrive at the condition $U(X_I,iY_I)=0$, where we also took into account that the energy along the trajectory has to vanish, since it is conserved by the dynamics \cite{Gorokhov98}. Then one finds that $Y_I =\mc{J}(X_I)$, which takes the following simplified form for the expanded potential \eqref{ExpandPot},
\begin{align}
\mc{J}(X_I) =  \sqrt{X_I \frac{\Vmax^x Y_d^3 (X_I-X_d)}{\Vmax^y X_d^3(X_I-Y_d)} } \,. 
\end{align}
The $X_I$ variable ranges from zero up to the turning point $X_d$, calculated by the requirement $\mc{J}(X_d) = 0$, and $X_d = 2 a \sqrt{\epsilon}$ for the expanded potential $\bar{U}$  \eqref{ExpandPot}. The instanton trajectories, defined by the equipotential lines $U(X_I, i Y_I)=0$, are illustrated in Fig.\ref{fig:Merge}-(a), for $\lambda = 1.86 $, $\lambda_d = 2.5 \lambda$, and for $\epsilon = 0.096$, with $F_c= 0.13$. Fig.\ref{fig:Merge}-(b) compares the values of $X_d$, derived by both the potential $U$ and $\bar{U}$ as a function of $\epsilon$, and implies that $\bar{U}$ is a good approximation of $U$, as long as $\epsilon \lesssim 0.05$. 

The quantum tunneling of the particle into the classically forbidden region is achieved by the nontrivial instanton solution $ (X_I,Y_I)$, in which the skyrmion starts at  $X=0$ at $\tau= -\infty$, reaches $X=X_d$ at $\tau=0$, and then returns to $X= 0$ at $\tau = \infty$. This motion occurs with a characteristic tunnel frequency 
\begin{align}
\omega_{\tau}=\frac{\sqrt{\Vmax^x \Vmax^y}}{\vert \tilde{Q} \vert X_d Y_d} = \frac{9 V_0  (3 \epsilon)^{1/4} (\alpha d)^2}{16 \hbar \vert \tilde{Q} \vert a^4}\,,
\label{TunFreq}
\end{align}
where for the rest of the paper units are restored. The probability of tunneling is governed by the temperature-independent WKB exponent, $e^{-\mc{S}_0}$, with the tunneling action $\mc{S}_0 = \mc{S}[X_I,Y_I]$ given by
\begin{align}
\mc{S}_0=2 \hbar \vert \tilde{Q} \vert  \int_0^{X_d} dX_I  [\mc{J}(X_I)-\mc{J}'(X_I) X_I ]  \simeq  \frac{16 \Vmax^x J_0}{15 \omega_\tau}\,,
\label{TunnAction}
\end{align}
where for the last approximate equality we used the expanded potential \eqref{ExpandPot}, and is further simplified as $\mc{S}_0 \simeq 5.6 \hbar \vert \tilde{Q} \vert a^2 \epsilon^{5/4}/(\alpha d)^2$. We note that the tunneling action depends on the width of the pinning potential $a$, but is independent of its height $V_0$ \cite{Lin13b}, and the coercive force $F_c$, in contrast to the tunneling exponent of domain walls in ferromagnets \cite{Braun97}. The dependence of $\mc{S}_0$ on $\epsilon$ is depicted in Fig.~\ref{fig:Merge}-(c), for a skyrmion with radius $\lambda = 1.86 \alpha d$, while the inset summarizes the dependence of $\mc{S}_0$ from the skyrmion size $\lambda$. The decay rate $\Gamma$ at zero temperature is calculated as \cite{Grabert84},
\begin{align}
\Gamma \simeq \frac{\omega_\tau}{2\pi}e^{-\mc{S}_0/\hbar} \simeq \frac{9 V_0 (3 \epsilon)^{1/4}  (\alpha d)^2}{32 \pi \hbar \vert \tilde{Q} \vert a^4}e^{-\mc{S}_0/\hbar} \,,
\label{TunRate}
\end{align}
with a characteristic  dependence on $\epsilon$ in the exponent $\propto \epsilon^{5/4}$ and the prefactor $\propto \epsilon^{1/4}$
that provide experimental signatures of quantum tunneling. To make quantum effects observable, two conditions must be satisfied. First, the inverse escape rate $\Gamma^{-1}$ must not exceed a few hours \cite{Braun97}, and second,  the thermal activation events over the barrier do not dominate over the quantum tunneling-induced transitions. The decay rate becomes determined solely by quantum effects below a characteristic temperature $T_c= \hbar U_0 /k_B \mc{S}_0 = 5 \hbar\omega_\tau /36 k_B$ \cite{Henggi90}, where $U_0=4 \Vmax^x/27$ is the height of the potential barrier. 
\begin{table*}
\caption{\label{Units} Tunneling quantities for the chiral magnetic insulator Cu$_2$OSeO$_3$, with $J_0=3.34$ meV, $D=0.79$ meV, $K=6.8\times 10^{-2}$ meV, $M_s=111.348$ kA m$^{-1}$, $\alpha=8.911$ \AA, $S=M_s \alpha^3/g\mu_B$ \cite{Janson2014}, and $Q=1$, $\lambda_d=\lambda$, $J'/J_0=0.3$, $D'=0$, and $N_A =30$.}
\begin{ruledtabular}
\begin{tabular}{ccccccccc}
 $\lambda$ &N&$\epsilon$&$X_d$ &$\omega_\tau$&$\mc{S}_0/\hbar$&$\Gamma^{-1}$&$T_c$
\\ \hline
\\
  $4.3$ nm& $8.8\times10^{2}$&$5\times 10^{-2}$ &3.02 nm&$2.90 \times 10^{10}$ s$^{-1}$&288.76& $5.51\times 10^{115}$ s&30.76 mK   \\
   & & $2\times 10^{-3}$&0.60 nm&$1.30 \times 10^{10}$ s$^{-1}$&5.16& $8.48 \times 10^{-8}$ s& 13.76 mK   \\
  & & $5 \times10^{-4}$&0.30 nm&$9.17 \times 10^9$ s$^{-1}$&0.91 &$1.71\times 10^{-9}$ s&  9.73 mK  \\
 \hline
 \\
  $7.4$ nm& $2.61\times10^{3}$&$5\times 10^{-2}$ &5.3 nm&$3.54 \times 10^9$ s$^{-1}$&886.59& $1.96 \times 10^{376}$ s&3.76 mK   \\
   & & $2 \times 10^{-3}$&1.06 nm&$1.58 \times 10^9$ s$^{-1}$&15.86& $0.03$ s& 1.68 mK   \\
  & & $5\times10^{-4}$&0.5 nm&$1.12 \times 10^9$ s$^{-1}$&2.80& $9.25\times 10^{-8}$ s&  1.19 mK  \\
   \hline
 \\
  $10.3$ nm& $5.05\times10^{3}$&$5\times 10^{-2}$ &7.40 nm&$1.04 \times 10^9$ s$^{-1}$&1731.46& $5.56 \times 10^{743}$ s&1.10 mK   \\
   & & $2 \times10^{-3}$&1.48 nm&$4.66 \times 10^8$ s$^{-1}$&30.97 & $38.15 \times 10^{4}$ s& 0.49 mK   \\
  & & $5\times10^{-4}$&0.74 nm&$3.29 \times 10^8$ s$^{-1}$&5.47& $4.55\times 10^{-6}$ s&  0.35 mK  \\
\end{tabular}
\end{ruledtabular} 
\label{Table}
\end{table*}
Table~\ref{Table} summarizes typical values of the tunneling exponent $\mc{S}_0$, the oscillation frequency $\omega_\tau$, the inverse tunneling rate $\Gamma^{-1}$, and the crossover temperature $T_c$, for various skyrmion radii and values of $\epsilon$, for the chiral magnetic insulator Cu$_2$OSeO$_3$, which is known to support stable skyrmions \cite{Seki2012}. For sufficiently small skyrmions with a radius of a few lattice sites, coherent tunneling out of a pinning potential is expected to take place involving some thousands of spins, within a few seconds, in the millikelvin temperature regime. To optimize the observability of quantum tunneling events, there is some flexibility to increase $\epsilon$ in this material, by reducing its thickness. For $N_A=5$ and $\epsilon=0.04$ we find   $\Gamma^{-1}=136$ s and $T_c=203$ mK.

The action \eqref{EuclAction} assumes a quasi-2D behavior, established when the transverse degrees of freedom are frozen out due to the finite number of layers $N_A$. The transverse magnon excitations of a bulk sample, with energy $\omega(k_z) = \mc{A}(k_z + l_0^{-1})^2 + g\mu_B H + 2 K -\mc{A} l_0^{-2}$ \cite{Garst2017}, where $l_0 = 2 \alpha d$, acquire an additional finite size gap which arises since $k_z^{\mbox{\tiny{min}}}= \pi / w$,  with $w = N_A \alpha$. We introduce $\mc{A} = 2 J_0 g \mu_B/\alpha M_s$, with $M_s$ the saturation magnetization. All transverse excitations freeze out below a critical width given by $w(T) =\pi g/(k_B T - g \mu_B H - 2K)$, with $g= \mc{A}l_0^{-1} + (\mc{A} k_B T +\mc{A}^2 l_0^{-2} + \mc{A} [g \mu_ B H + 2 K])^{1/2} $. To make an estimate for Cu$_2$OSeO$_3$, we use the parameters summarized in Table.~\ref{Table}, and a choice of $H= 345.6$ mT and $\lambda= 8.31 \alpha$. For a freezing temperature of $T = 2.2$ K, we find $N^{\mbox{\scriptsize {max}}}_A= w(T)/\alpha = 114$ layers. 

In the special case of a separable pinning potential $U(X,Y)= U_1(X)+U_2(Y)$, the problem can be reduced to a one-dimensional massive particle by integrating out the $Y$ variable from Eq.~\eqref{LagrRT} \cite{Zarzuela13}, while the properties Eqs.\eqref{TunFreq}-\eqref{TunRate} remain unaffected.  Environmental degrees of freedom are a source of dissipation and could suppress the probability of the tunneling process. Ohmic couplings have the most detrimental effect on the tunneling rate of mesoscopic systems \cite{Caldeira83}, while super-Ohmic interactions have a very small contribution \cite{Chudnovsky92}. In insulators and for low temperatures below the magnon gap, which is about $10$ K for the parameters of Table~\ref{Table}, the super-Ohmic skyrmion-magnon interaction is the main source of dissipation \cite{Psaroudaki18}. In the presence of a pinning potential, the lowest Landau level splits into quantized levels with spacing proportional to the potential height \cite{Lin13}. Quantum tunneling processes that involve the excitation of the skyrmion from the LLL to the next higher one, mediated by thermal excitations, are expected to provide a better estimation of $T_c$ \cite{Leuenberger2000}. Such processes require a detailed understanding of the rates of the thermal excitations, and we thus leave it for future work. In view of the increasing interest on new insulating materials that enable the stabilization of skyrmions \cite{Ding19}, we anticipate that our results will initiate experimental studies towards the possibility of observing a quantum mechanical behavior at a mesoscopic scale for a topological particle. 

\begin{acknowledgments} This work was supported by the Swiss National Science Foundation (Switzerland) and the NCCR QSIT.
\end{acknowledgments}

\end{document}